\documentclass[a4paper]{jpconf}

\usepackage{graphics}
\usepackage{graphicx}
\usepackage{epsfig}
\usepackage{units}
\usepackage{cite}
\usepackage[switch]{lineno}
\usepackage{calc}
\usepackage{endnotes}

\newcommand{\pip}{$\pi^{+}$}
\newcommand{\pim}{$\pi^{-}$}
\newcommand{\kap}{K$^{+}$}
\newcommand{\kam}{K$^{-}$}
\newcommand{\pbar}{$\rm\overline{p}$}

\newcommand{\s}{$\sqrt{s}$}
\newcommand{\dedx}{d$E$/d$x$}

\newcommand{\pp}{pp}

\newcommand{\kos}{{K$^{0}_{S}$}}
\newcommand{\lam}{$\Lambda$}
\newcommand{\lambar}{$\rm\overline{\Lambda}$}
\newcommand{\xip}{{$\Xi^{+}$}}
\newcommand{\xim}{{$\Xi^{-}$}}
\newcommand {\mmass} {\mbox{\rm MeV$\kern-0.15em /\kern-0.12em c^2$}}

\begin{document}
\title{Measurement of identified charged hadron spectra
 with the ALICE experiment at the LHC}

\author{Leonardo Milano}

\address{Dipartimento di Fisica Sperimentale dell'Universit\'a and Sezione INFN, Turin, Italy}

\ead{milano@to.infn.it}

\begin{abstract}
The ALICE experiment features multiple particle identification systems.
The measurement of the identified charged hadron $p_{t}$ spectra in proton-proton collisions at $\sqrt{s}=900$ GeV  
will be discussed.
In the central rapidity region ($|\eta|<0.9$) particle identification and tracking are performed using the Inner Tracking System (ITS), which is the closest detector to the beam axis, the Time Projection Chamber (TPC) and a dedicated time-of-flight system (TOF).
Particles are mainly identified using the energy loss signal in the ITS and TPC.
In addition, the information from TOF is used to identify hadrons at higher momenta.
Finally, the kink topology of the weak decay of charged kaons provides an alternative method to extract the transverse momentum spectra of charged kaons.
This combination allows to track and identify charged hadrons in the transverse momentum ($p_{t}$) range from 100~MeV/c up to 2.5 GeV/$c$.
Mesons containing strange quarks (\kos, $\phi$) and both singly and doubly strange baryons (\lam, \lambar, and \xip + \xim) are identified by their decay topology inside the TPC detector.
Results obtained with the various identification tools above described and a comparison with theoretical models and previously published data will be presented. 
\end{abstract}

\section{Introduction}


The transverse momentum spectra and yields of
identified particles at mid-rapidity from the first \pp\ collisions are presented. Data are collected in the autumn of 2009, during the commissioning of the
LHC,  at \s~= 900 GeV. The evolution of particle production in
\pp\ collisions with collision energy is studied by comparing data from previous experiments.

The \pip, \pim, \kap, \kam, p, and \pbar\ distributions,
identified via several independent techniques utilizing specific
energy loss, \dedx, information from the Inner Tracking System
(ITS) and the Time Projection Chamber (TPC), velocity
measurements in the Time-Of-Flight array (TOF) are reported. The combination of
these methods provides particle identification over the transverse
momentum range 0.1 GeV/c~ $<\ensuremath{p_{\rm t}} <$~ 2.5 GeV/c.
Charged kaons, identified via kink topology of their weak decays
in the TPC, provide a complementary measurement over a similar
\ensuremath{p_{\rm t}} range. All reported particle yields are for primary
particles, namely those directly produced in the collision
including the products of strong and electromagnetic decays but
excluding weak decays of strange particles.


\section{Experimental setup and data analysis}
\label{exp_data}

\subsection{The ALICE detector}
\label{experiment}

The ALICE detector and its performance are described in
detail in \cite{Aamodt:2008zz}.
For the analyses described the following detectors
are used: the ITS, the TPC and the TOF detector. These detectors
are positioned in a solenoidal magnetic field of $B$~=~0.5~T and
have a common pseudo-rapidity coverage of $-0.9 < \eta < 0.9$. Two
forward scintillator hodoscopes (VZERO) are used for triggering
purposes. They are placed on either side of the interaction
region, covering regions $2.8 < \eta < 5.1$ and $-3.7 < \eta <
-1.7$.
\begin{itemize}
\item{The Inner Tracking System -}
The ITS is the closest of the central barrel detectors to the beam
axis. It is composed of six cylindrical layers of silicon
detectors. The two innermost layers are equipped with pixel
detectors (SPD), followed by two layers of drift detectors (SDD)
and two layers of double-sided silicon strip detectors (SSD). The
innermost layer is at 3.9 cm from the beam axis, while the outer
layer is at 43.0 cm.
The four layers equipped with SDD and SSD also provide
a measurement of the specific energy loss \dedx . The ITS is also
used as a standalone tracker to reconstruct charged particles
with momenta below 200 MeV/$c$ that are deflected or decay before
reaching the TPC, and to recover tracks crossing dead regions of
the TPC.

\item{The Time Projection Chamber -}
The TPC is  the main tracking device. It is a large volume, high
granularity, cylindrical detector with an outer radius of 2.78 m
and a length of 5.1 m. The active volume extends from 0.85 m to
2.47 m in radius. It covers 2$\pi$ in azimuth and $|\eta|<0.9$ in
polar angle for the full radial track length. Accepting  one third
of the full radial track length extends the range  to $|\eta|<$
1.5. The 90 m$^3$ drift volume is filled with a Ne (85.7\%),
CO$_2$ (9.5\%), and N$_2$ (4.8\%) gas mixture. A high voltage
central membrane splits the drift region in two halves, resulting
in a maximal drift time of 94 $\mu$s. The TPC has a good particle identification capability thanks to the measurement of energy loss of particles in the gas mixture.

\item{The Time-Of-Flight Detector -}
The TOF detector consists of 18 azimuthal sectors, each containing
91 Multi-gap Resistive Plate Chambers (MRPCs) distributed in five
gas-tight modules. It is positioned at 370-399 cm from the beam
axis. The region $\rm{260^\circ}$ $< \phi $ $< \rm{320^{\circ}}$
at $\eta \sim 0$ is not covered in order to minimize the material
in front of the Photon Spectrometer, which is not used in this
analysis. The MRPC detectors are installed with a projective
geometry along the beam direction, minimizing the variation of the
flight path of particles across the sensitive area of the
detector. Each MRPC is segmented into 96 read-out pads (2.5
$\times$ 3.5 $\rm{cm^2}$ size).
\end{itemize}

\subsection{Event selection and normalization}
\label{events}

The data presented were collected during the
commissioning of the LHC at CERN in the autumn of 2009, with \pp\
collisions at $\sqrt{s}=900$ GeV. The collider was run with four
bunches per beam, resulting in two bunch crossings per beam
circulation period (89~$\mu$s) at the ALICE interaction point. The
remaining two bunches per beam were not collided at ALICE, and
served to estimate the contribution of beam-gas interactions. The
average event rate was a few Hz, so the fraction of pile-up events
was negligible.

\section{\pip, \pim, \kap, \kam, p, and \pbar\ identification in ALICE}
The \dedx{} and TOF signals are used for particle identification
as a function of the momentum $p$, whereas the final spectra are
given as a function of the transverse momentum \ensuremath{p_{\rm t}}.

In the case of the TPC and ITS analyses, particles were identified via
the specific energy loss \dedx. Unique identification on a
track-by-track basis is possible in regions of momentum where the
bands are clearly separated from each other. In overlapping areas,
particle identification is still possible on a statistical basis using
fits to the energy loss distribution in each \ensuremath{p_{\rm t}}-bin. The fits are
performed on the distribution of the difference between the measured
and the expected energy deposition for tracks within the selected
rapidity range $|y|$ $<$ 0.5. 
The calculated expected energy loss depends on the
measured track momentum $p$ and the assumed mass for the particle. The
procedure is therefore repeated three times for the entire set of
tracks, assuming the pion, kaon, and proton mass.

In the TPC analysis, the difference [\dedx]$_{\rm meas} -
$[\dedx$(pid, p_{tot})]_{\rm calc}$ is used. For the ITS the
difference of the logarithm of the measured and calculated energy
deposit $\ln[$\dedx$_{\rm
meas}]-\ln[$\dedx($pid$,$p_{tot}\rm)_{\rm calc}]$ is taken to
suppress the non-Gaussian tails originating from the smaller
number of \dedx\ measurements.

In the case of the TOF, the identification is based on the
time-of-flight information. The procedure for the extraction of
the raw yields differs slightly from the one used for TPC and ITS.

\subsection{Particle identification in the ITS}
\label{ITS}
In both the ITS standalone (track reconstruction only using the ITS) and in the ITS-TPC (reconstruction is performed using both the ITS and the TPC) analyses, the
\dedx\ measurement from the SDD and the SSD is used to identify
particles. The standalone tracking result extends the momentum
range to lower \ensuremath{p_{\rm t}} than can be measured in the TPC, while the
combined tracking provides a better momentum resolution.

For each track, \dedx\ is calculated using a truncated mean: the average of
the lowest two points in case four points are measured, or a
weighted sum of the lowest (weight 1) and the second lowest point
(weight 1/2), in case only three points are measured.


\begin{figure}[h]
\includegraphics[width=16pc]{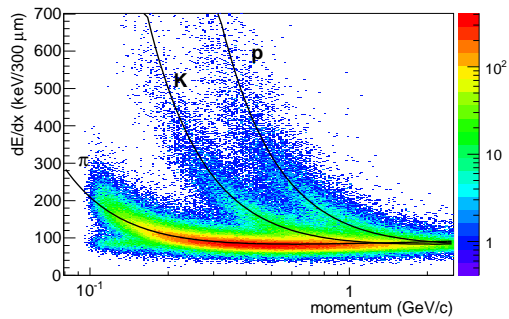}\hspace{2pc}%
\begin{minipage}[b]{16pc}\caption{\label{fig:ITSsadedx}Specific energy loss \dedx\ vs.~momentum in pp collisions at $\sqrt{s}=$~0.9  TeV 
for ITS standalone tracks measured with the ITS. The solid lines are a
parametrization (from \cite{Back:2006tt}) of the detector response
based on the Bethe-Bloch formula.}
\end{minipage}
\end{figure}

Figure~\ref{fig:ITSsadedx} shows the truncated mean \dedx\ for the
sample of ITS standalone tracks along with the PHOBOS
parametrization of the most probable value~\cite{Back:2006tt}.

For the ITS standalone track sample, the histograms are fitted
with three Gaussians and the integral of the Gaussian centered at
zero is used as the raw yield of the corresponding hadron species.

For the ITS-TPC combined track sample, a non-Gaussian tail is
visible. This tail is a remnant of the tail of the Landau
distribution for energy loss. It was verified using simulations
that the shape and size of the tail are compatible with the
expectations for a truncated mean using two out of four samples.

\begin{figure}[h]
\begin{minipage}[b]{16pc}\includegraphics[width=15pc]{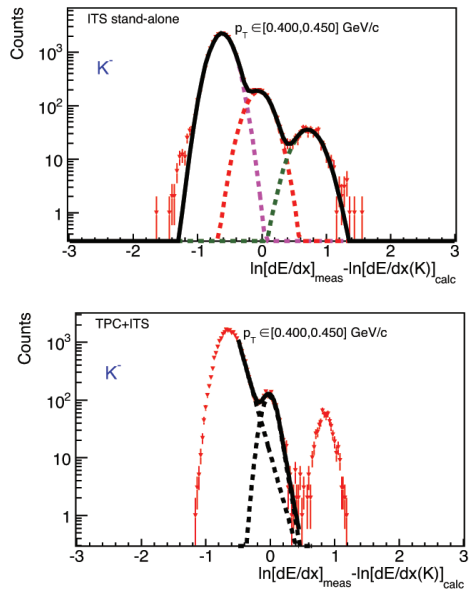}\hspace{2pc}%
\end{minipage}\hspace{2pc}
\begin{minipage}[b]{16pc}\caption{\label{its:3gausITS}Distribution of ln[\dedx]$_{\rm
meas}-$ln[\dedx({\rm K})]$_{\rm calc}$ measured with the ITS in
the \ensuremath{p_{\rm t}}-range 400-450
MeV/$c$, using the kaon mass hypothesis. The upper
panel shows the result for ITS standalone tracks, while the
lower panel shows the ITS-TPC combined result.  The lines indicate
fits as described in the text.}
\end{minipage}
\end{figure}

Examples of \dedx\
distributions are shown in Fig.~\ref{its:3gausITS} for negative
tracks using the kaon mass hypothesis in a given \ensuremath{p_{\rm t}}
interval ( 400-450
MeV/$c$) for both ITS standalone tracks (upper panel) and
ITS-TPC combined tracks (lower panel).

\paragraph{Efficiency correction}

The raw hadron yields extracted from the fits to the \dedx\
distributions are corrected for the reconstruction efficiency
determined from Monte Carlo simulations, applying the same
analysis criteria to the simulated events  as to the data.
Secondary particles from interactions in the detector material and
strange particle decays have been subtracted from the yield of
both simulated and real data. The fraction of secondaries after
applying the track impact-parameter cut depends on the hadron
species and amounts
 to 1-3\% for pions and 5-10\% for protons depending on \ensuremath{p_{\rm t}}. The
secondary-to-primary ratio has been estimated by fitting the
measured track impact-parameter distributions with three
components, prompt particles, secondaries from strange particle
decays and secondaries produced in the detector material for each
hadron species. Alternatively, the contamination from secondaries
have been determined using Monte Carlo samples, after rescaling
the $\Lambda$ yield to the measured values~\cite{strange}. The
difference between these two procedures is about 3\% for protons
and is negligible for other particles.

\subsection{Particle identification in the TPC}
\label{TPC}

\begin{figure}[h]
\includegraphics[width=16pc]{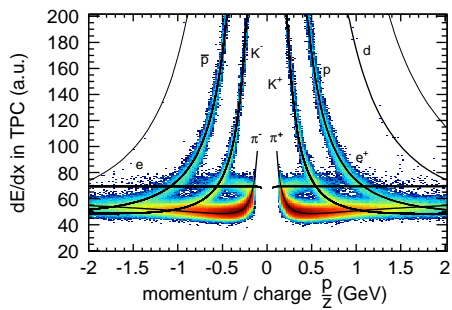}\hspace{2pc}%
\begin{minipage}[b]{16pc}\caption{\label{tpc:dedx}Specific energy loss \dedx\ vs.~momentum in pp collisions at $\sqrt{s}=$~0.9  TeV 
for tracks measured with the ALICE TPC. The solid lines are a
parametrization of the Bethe-Bloch curve~\cite{ALEPH}. }
\end{minipage}
\end{figure}

Particle identification is based on the specific energy deposit of
each particle in the drift gas of the TPC (up to 159 independent measurements for tracks crossing the whole TPC volume), shown in Fig.~\ref{tpc:dedx} as a function of momentum for positive and negative charges.   The solid curves show the
calibration curves obtained by fitting the ALEPH parametrization
of the Bethe-Bloch curve~\cite{ALEPH} to the data points in
regions of clear separation.

As in the case of the ITS, a truncated-mean procedure is used to
determine \dedx\ (60\% of the points are kept).  This reduces the
Landau tail of the \dedx{} distribution to the extent that it is
very close to a Gaussian distribution.

\begin{figure}[h]
\includegraphics[width=16pc]{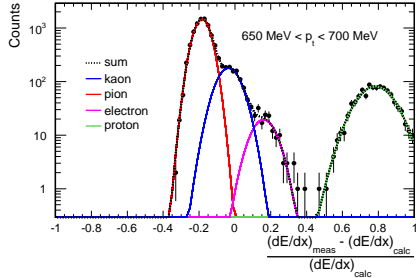}\hspace{2pc}%
\begin{minipage}[b]{16pc}\caption{\label{tpc:pid}Distribution of ([\dedx]$_{\rm
meas} - $[\dedx(kaon)]$_{\rm calc}$) / [\dedx(kaon)]$_{\rm calc}$
measured with the TPC in a given \ensuremath{p_{\rm t}} -bin showing the separation
power. The solid lines are Gaussian fits to the distributions.}
\end{minipage}
\end{figure}

One example of the \dedx\ distribution in a specific \ensuremath{p_{\rm t}} bin in the kaon mass hypothesis is shown in
Fig.~\ref{tpc:pid}. The peak centered at zero refers to kaons and
the other peaks are from other particle species. As the background
in all momentum bins is negligible, the integrals of the Gaussian
give the raw yields.

\paragraph{Efficiency correction}

The raw hadron spectra are corrected for the reconstruction
efficiency, determined by doing
the same analysis on MonteCarlo events. The efficiency is
calculated by comparing the number of reconstructed particles to
the number of charged primary particles from PYTHIA in the chosen
rapidity range.  
The range with a reconstruction efficiency lower than 60\% (for
pions and protons) is omitted for the analysis corresponding to a
low-\ensuremath{p_{\rm t}} cut-off of 200 MeV/$c$ for pions, 250 MeV/$c$ for kaons,
and 400 MeV/$c$ for protons.

Protons are corrected for the contamination of secondaries from
material and of feed down from weak decays. The feed down was
determined by two independent methods as described in ~\ref{ITS}. 

\subsection{Particle Identification with the Time Of Flight detector}
\label{sec:TOF}

Particles reaching the TOF system are identified by measuring
simultaneously their momentum and velocity.
The flight path of the track $L$ is defined as the distance along
the track trajectory between the point of its closest approach to
the event vertex and the TOF sensitive surface.
The time of flight is calculated for different mass hypotheses by
summing up, at each tracking step, the time-of-flight increments
$\Delta t_k = \Delta l_k \sqrt{p^2_k+m_i^2}/p_k$, with $p_k$ being
the local value of the track momentum, and $\Delta l_k$ the
track-length increment along its trajectory.

\begin{figure}[h]
\includegraphics[width=16pc]{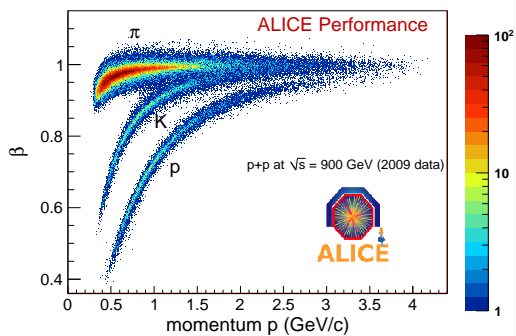}\hspace{2pc}%
\begin{minipage}[b]{16pc}\caption{\label{TOF:beta_vs_p}$\beta$ vs. momentum of the tracks in the TOF.}
\end{minipage}
\end{figure}

From the reconstructed flight path $L$ and the measured time of
flight $t_{\rm{TOF}}$, the velocity $\beta= L/t_{\rm TOF}$ is
obtained (as displayed in Fig.~\ref{TOF:beta_vs_p}) as a function
of the momentum $p$ at the vertex. Clearly visible are the bands
corresponding to charged pions, kaons and protons. The width of
the bands reflects the observed overall time-of-flight resolution
of about 180 ps, which has contributions from the TOF time
resolution, the accuracy of the reconstructed flight path and
the uncertainty on the event start time, $t_{0}^{ev}$.

The yields of $\pi$, K and p are obtained from the simultaneous
fit of the distribution of the time difference $S$ between
measured $t_{\rm TOF}$ and the average between the calculated time
averaging for pions and kaons
\begin{equation}
 S = t_{\rm TOF}-(t_{\rm calc}^{\pi} + t_{\rm calc}^{\rm K})/2.
\end{equation}
The choice of this variable is motivated by the observation that
the identification procedure is optimized when pions and kaons
have a symmetric treatment. Moreover, extracting the yield for
different species in a simultaneous fit guarantees that the
resulting number of pions, kaons and protons matches the total
number of tracks in the given momentum bin.

\begin{figure}[h]
\includegraphics[width=16pc]{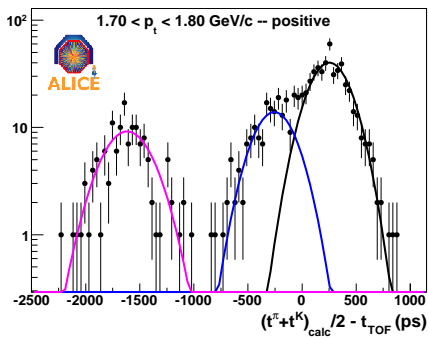}\hspace{2pc}%
\begin{minipage}[b]{16pc}\caption{\label{TOF:pid}Distribution of the time difference between the TOF
signal and the average  of the expected times for pions and kaons in a given momentum bin for positively
charged particles. The fits are performed using Gaussian shapes.}
\end{minipage}
\end{figure}

The distribution of the variable ${S}$
is reported in Fig.~\ref{TOF:pid} for positive particles in a given transverse
momentum bin. 
The distribution shows a Gaussian behavior for each particle
species and the lines indicate the result of a three-Gaussian fit
used to extract the raw yields.

\paragraph{Efficiency Correction}

As the track selection used in the TOF analysis is the same as the
one described in the TPC analysis (subsection~\ref{TPC}), the same
tracking corrections are applied. In the case of TOF analysis, an
additional correction is needed in order to take into account that
only a fraction of the particles reconstructed by the TPC are
associated to a signal in TOF. This matching efficiency includes
all sources of track losses in the propagation from the TPC to the
TOF (geometry, decays and interactions with the material) and its
matching with a TOF signal (the TOF intrinsic detector efficiency,
the effect of dead channels and the efficiency of the track-TOF
signal matching procedure).

\section{Particle identification in ALICE via topological criteria}
\subsection{Kaon Identification using their decay in the TPC}
\label{sec:kinks}
Charged kaons can also be identified by their weak decay (kink topology) inside the TPC
detector. These tracks are rejected in the previously
described TPC analysis.
The decay
channels with the highest branching ratio (B.R.) for kaons are
the two-body decays
\begin{itemize}
\item [(1)] {$\rm{K^\pm}\rightarrow\mu^\pm +\nu_{\mu}$} ,  (B.R.
63.55\%) \item [(2)] {$\rm{K^\pm}\rightarrow\pi^\pm + \pi^0$}  ,
(B.R. 20.66\%).
\end{itemize}

Three-body decays with one charged daughter track (B.R. 9.87\%) as
well as three-body decays into three char\-ged pions (B.R. 5.6\%)
are also detected.

\begin{figure}[h]
\includegraphics[width=15pc]{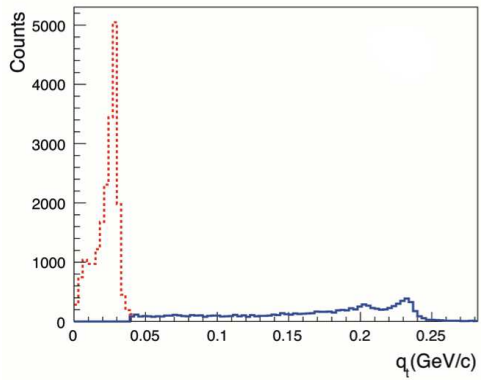}\hspace{2pc}%
\begin{minipage}[b]{16pc}\caption{\label{fig:KinkSelection}$q_{\rm t}$ distribution in pp collisions at $\sqrt{s}=$~0.9  TeV  of the
daughter tracks with respect to mother momentum for
reconstructed kinks inside the analyzed sample. The dashed (solid)
histograms show the distribution before (after) applying the
$q_{\rm t}
> 40$~MeV/$c$ cut.
}
\end{minipage}
\end{figure}

The identification of  kaons from kink topology and its separation
from pion decay is based on the decay kinematics. The transverse
momentum of the daughter with respect to the mother's direction,
$q_{\rm t}$, has  an upper limit of 236 MeV/$c$ for kaons and 30
MeV/$c$ for  pions for the two-body decay to $\mu +\nu_{\mu}$. The
corresponding upper limit for the two-body decay (2)
{$\rm{K^{\pm}}\rightarrow\pi^{\pm} +\pi^{0}$} is 205 MeV/$c$. All three limits
can be seen as the edges of the peaks in Fig.~\ref{fig:KinkSelection}, which
shows the $q_{\rm t}$ distribution of all measured kinks inside
the selected volume and rapidity range $|y|$ $<$ 0.7. Selecting
kinks with $q_{\rm t} > 40$ MeV/$c$ removes the majority of
$\pi$-decays as shown by the dashed (before) and solid (after)
histograms.

The invariant mass for the decay into $\mu^\pm +\nu_{\mu}$ is
calculated from the measured difference between the mother and
daughter momentum, their decay angle, assuming zero mass for the
neutrino.
%
%
%
  
\subsection{Topological reconstruction of \kos, \lam, \lambar}
\label{sec:v0}
  
  The measurement of \kos, \lam~ and  \lambar ~is based on the 
reconstruction of the secondary vertex (V0) associated to their weak decay.  
The V0 finding procedure starts with the selection of secondary tracks, i.e.
tracks having a sufficiently large impact parameter with respect to the primary vertex. 
All possible combinations between two secondary tracks of opposite charge 
are then examined.
  They are accepted as V0 candidates only if the DCA between them
is smaller than 0.5 cm. The minimization of the distance between the tracks is performed numerically
using helix parameterizations in 3D. 
The V0 vertex position is a point on the line connecting the points of closest approach
between the two tracks. Its distance from each daughter track is taken
to be proportional to the precision of the track parameter estimations.
Once their position is determined, only the V0 candidates located inside a 
given fiducial volume are accepted.
  
\subsection{Topological reconstruction of \xip and \xim}
\label{sec:cascade}
  The \xip~ and \xim ~particles are identified via their ``cascade'' decay topology.
The cascade finding procedure starts from the V0 finding procedure for the \lam(\lambar)
daughter but with less stringent selection criteria.
This is done to increase the efficiency and to allow for the fact  that the daughter
\lam's do not have to point back to the primary vertex.

The V0 candidates found within the \lam~mass window ($1116 \pm 6~\mmass$)
are combined with all possible secondary tracks (bachelor candidates) with the exception
of both V0 daughter tracks.
A cut on the impact parameter of the bachelor track is applied to reject the primary particles 
which increase the combinatorial background.

A V0-bachelor association is performed if the distance of closest approach between the
bachelor track and the V0 trajectory (DCA between V0 and bachelor track) is small (less than
$3 cm$).
Finally, this cascade candidate is selected if its reconstructed momentum points back to 
the primary vertex (cosine of cascade pointing angle).
The cascade finding is limited to the fiducial region used for V0 
reconstruction).

  \subsection{Reconstruction of $\phi$}
\label{sec:phi}
  The $\phi$ resonance is reconstructed through its principal decay channel
$\phi\rightarrow \rm{K}^+ \rm{K}^-$.
With a $c\tau$ of 45 fm, its decay vertex is indistinguishable from 
the primary collision vertex.
Therefore the selection criteria adopted for the candidate daughter tracks 
are the ones used for primaries.

A crucial issue for the $\phi$ reconstruction, as for any strongly decaying
resonance, is the combinatorial background determination.
In the present analysis PID is used to select kaons, rejecting most
of the background while leading to a very small loss in efficiency.
For this purpose, tracks are selected
if the PID information from the TPC is compatible with a kaon signal and
using the TOF signal when available.

  \subsection{Efficiency corrections}
\label{subsec_eff}
The efficiency corrections are obtained by analyzing Monte Carlo (MC) events in exactly
the same way as for the real events.
Little dependence is found on the several MC generators which are used.
Therefore the corrections presented here are obtained using
the event generator PYTHIA 6.4 (tune D6T)~\cite{Sjostrand:2006za} and 
GEANT3~\cite{Alice:Geant} for particle transport through the ALICE detectors.

\section{Results}
\label{sec:results}

\begin{figure}[h]
\begin{minipage}{16pc}
\includegraphics[width=15pc]{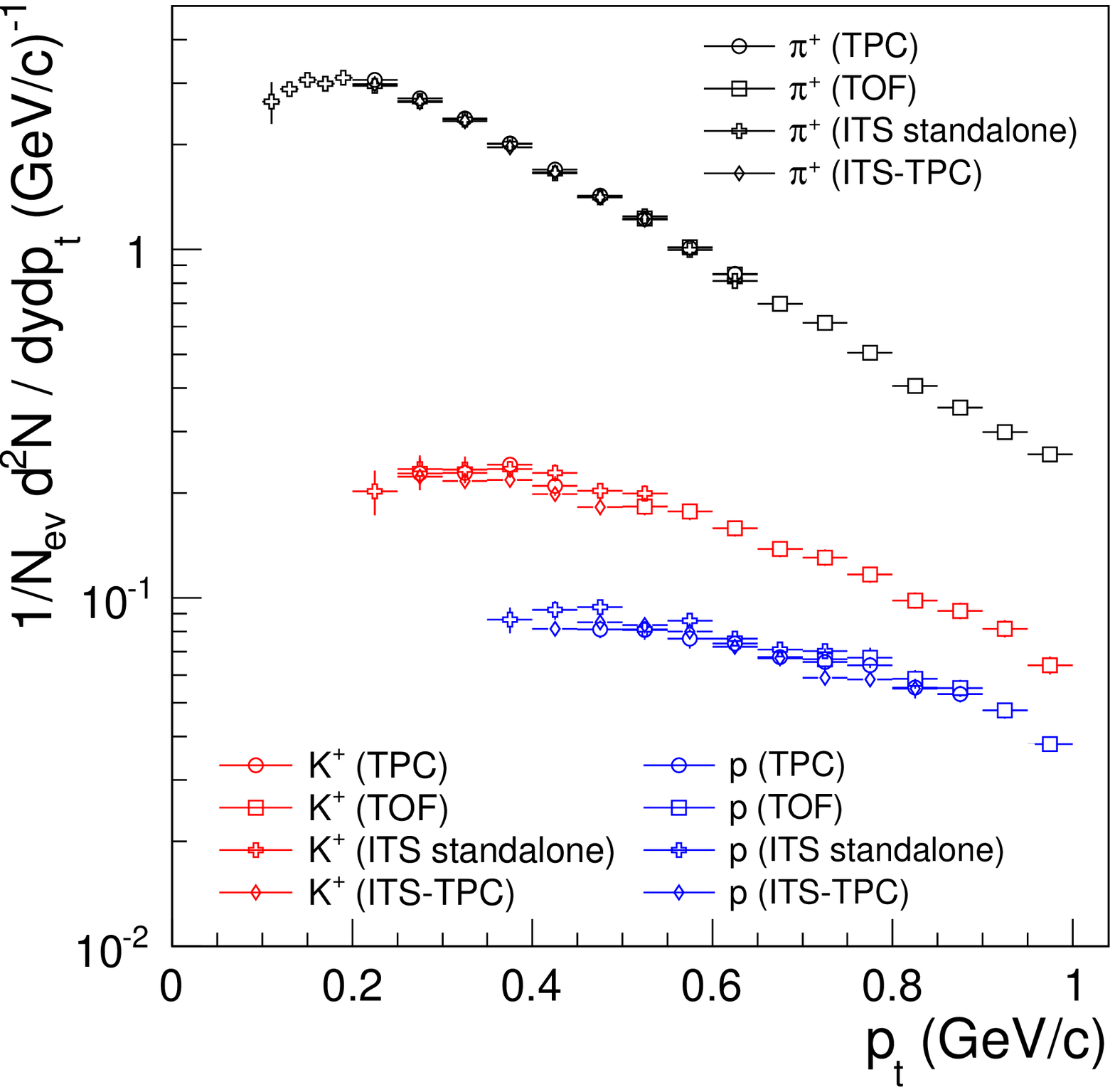}
\end{minipage}\hspace{2pc}%
\begin{minipage}{16pc}
\includegraphics[width=15pc]{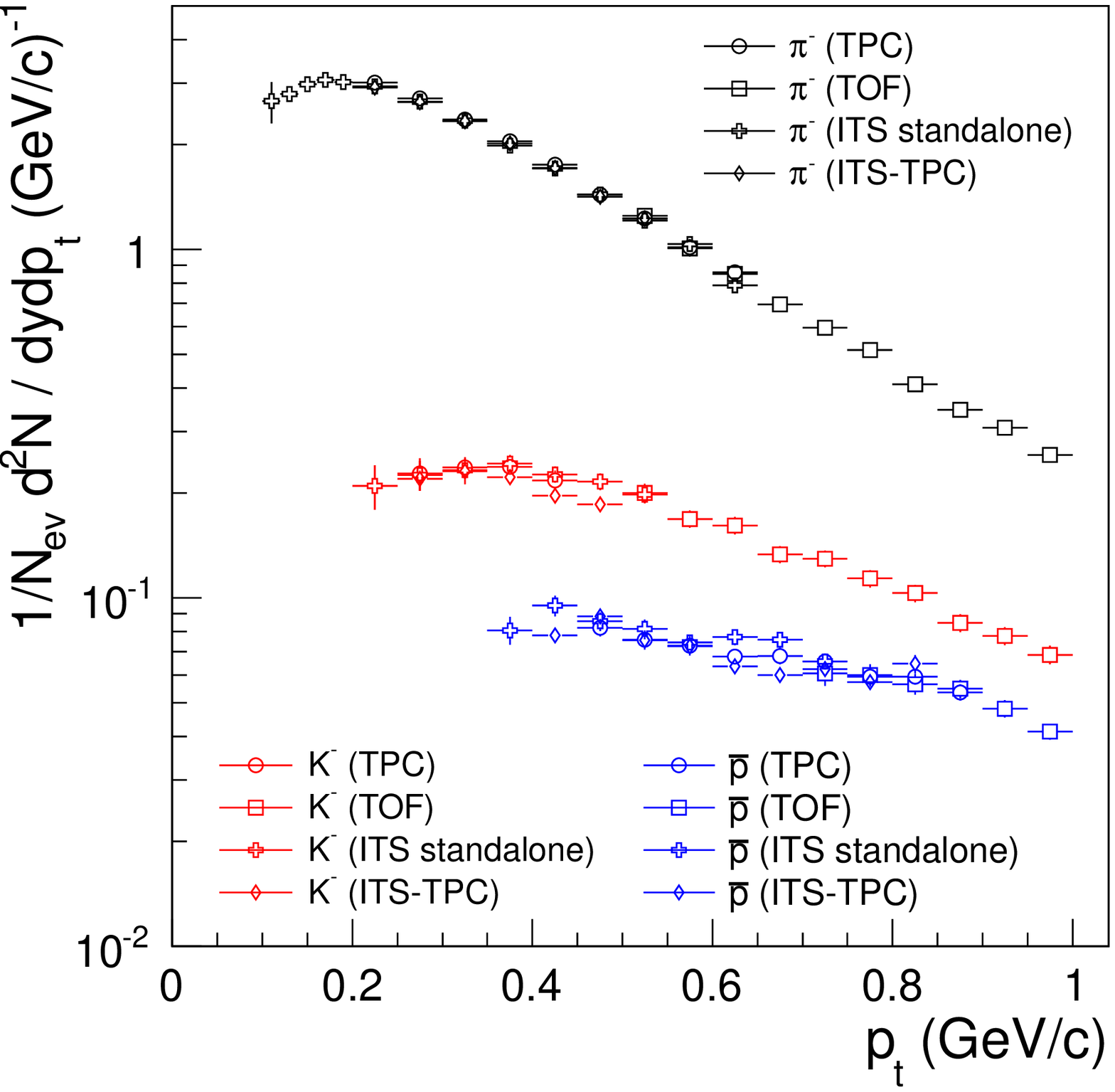}
\end{minipage}\hspace{2pc}%
\begin{minipage}{34pc}
\caption{\label{fig:spectra_det}Transverse momentum spectra ${\rm d}^2N / ({\rm d}p_{\rm t}{\rm d}y)$
 for $|y|<$ 0.5 of positive (left) negative (right) hadrons from the various analyses. Only systematic errors are plotted.}
\end{minipage} 
\end{figure}

\begin{figure}[h]
\begin{minipage}{16pc}
\includegraphics[width=15pc]{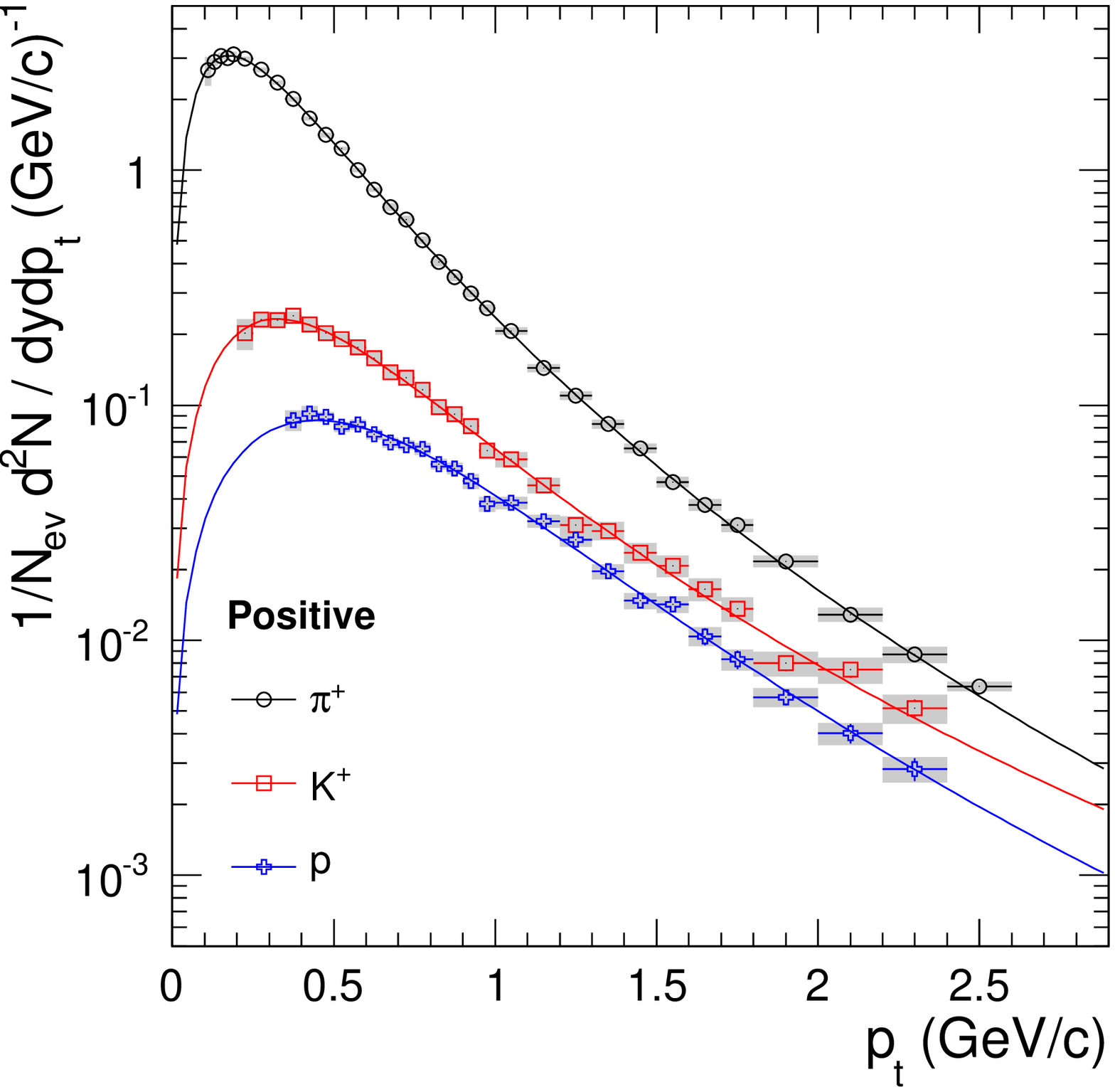}
\end{minipage}\hspace{2pc}%
\begin{minipage}{16pc}
\includegraphics[width=15pc]{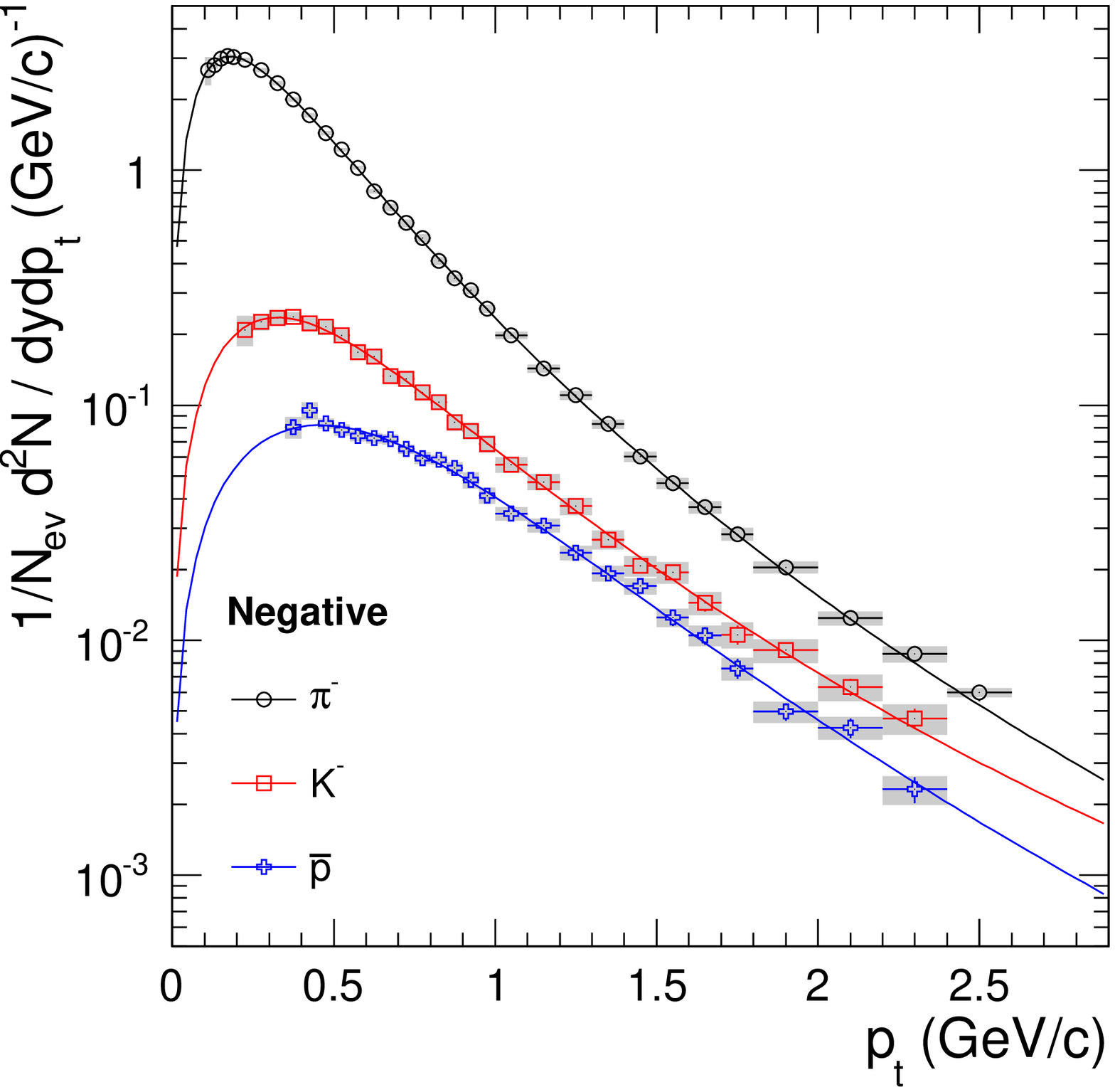}
\end{minipage}\hspace{2pc}%
\begin{minipage}{34pc}
\caption{\label{fig:spectra_detlevi}Transverse momentum spectra of positive (left) and negative (right) hadrons from pp collisions at \s = 900~
GeV. Grey bands: total \ensuremath{p_{\rm t}}-dependent error (systematic plus
statistical); normalization systematic error (3.6\%) not shown.
The curves represent fits using a L\'{e}vy function. }
\end{minipage} 
\end{figure}

Spectra from ITS standalone, ITS-TPC, TPC and TOF analyses are reported in Figure~\ref{fig:spectra_det}. 
The analyses from the
different detectors use a slightly different sample of tracks and
have largely independent systematics (mainly coming from the PID
method and the contamination from secondaries). The spectra have
been averaged, using the systematic errors as weights.  From this
weighted average, the combined, \ensuremath{p_{\rm t}}-dependent, systematic error is
derived. The combined spectra have an additional overall
normalization error, coming primarily from the uncertainty on the
material budget (3\%) and from the normalization procedure (2\%). Combined spectra are reported in Figure~\ref{fig:spectra_detlevi}. 

The combined spectra shown in Fig.~\ref{fig:spectra_detlevi} are fitted
with the L\'{e}vy (or Tsallis) function (see
e.g.~\cite{Tsallis:1987eu})
\begin{equation}
  \label{eq:1}
  \frac{{\rm d}^2N}{{\rm d}p_{\rm t}{\rm d}y} = p_{\rm t} \times \frac{{\rm d}N}
  {{\rm d}y} \frac{(n-1)(n-2)}{nC(nC + m_{0} (n-2))} \left( 1 + \frac{m_{\rm t} - m_{0}}{nC} \right)^{-n}
\end{equation}
  with the fit parameters $C$, $n$ and the yield ${\rm d}N/{\rm
  d}y$. This function gives a good description of the spectra and has
been used to extract the total yields and the $\langle p_{\rm t}
\rangle$.
  
\begin{figure}[h]
\includegraphics[width=16pc]{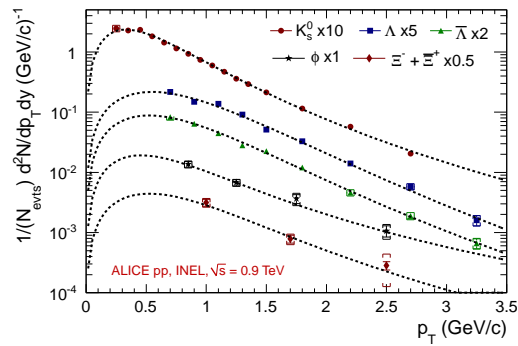}\hspace{2pc}%
\begin{minipage}[b]{16pc}\caption{\label{fig:corryield}Particle spectra (corrected yields) as a function of \ensuremath{p_{\rm t}} for \kos (circles),
\lam (squares), \lambar (triangles), $\phi$ (stars) and \xip + \xim
  (diamonds). The data points are scaled for visibility
and plotted at the centre of the bins.
The dotted curves show L\'{e}vy fits.
}
\end{minipage}
\end{figure}

The \ensuremath{p_{\rm t}} spectra for \kos, \lam, \lambar,  $\phi$ and \xip + \xim are shown
in Fig.~\ref{fig:corryield} along with the L\'{e}vy fits.
When comparing the different spectra, it is found that the inverse slope parameter
$T$ increases with the mass of the particle.
For example, it changes from $168 \pm 5$ ~MeV  for \kos~ to $229 \pm 15$~ MeV
for \lam ~when the L\'{e}vy  fit is used.
The \xip~ and \xim ~apparently do not follow this trend.
However, this is most likely because the very limited statistics do not allow for a
well-constrained fit.
  
%

\begin{figure}[h]
\begin{minipage}{16pc}
\includegraphics[width=15pc]{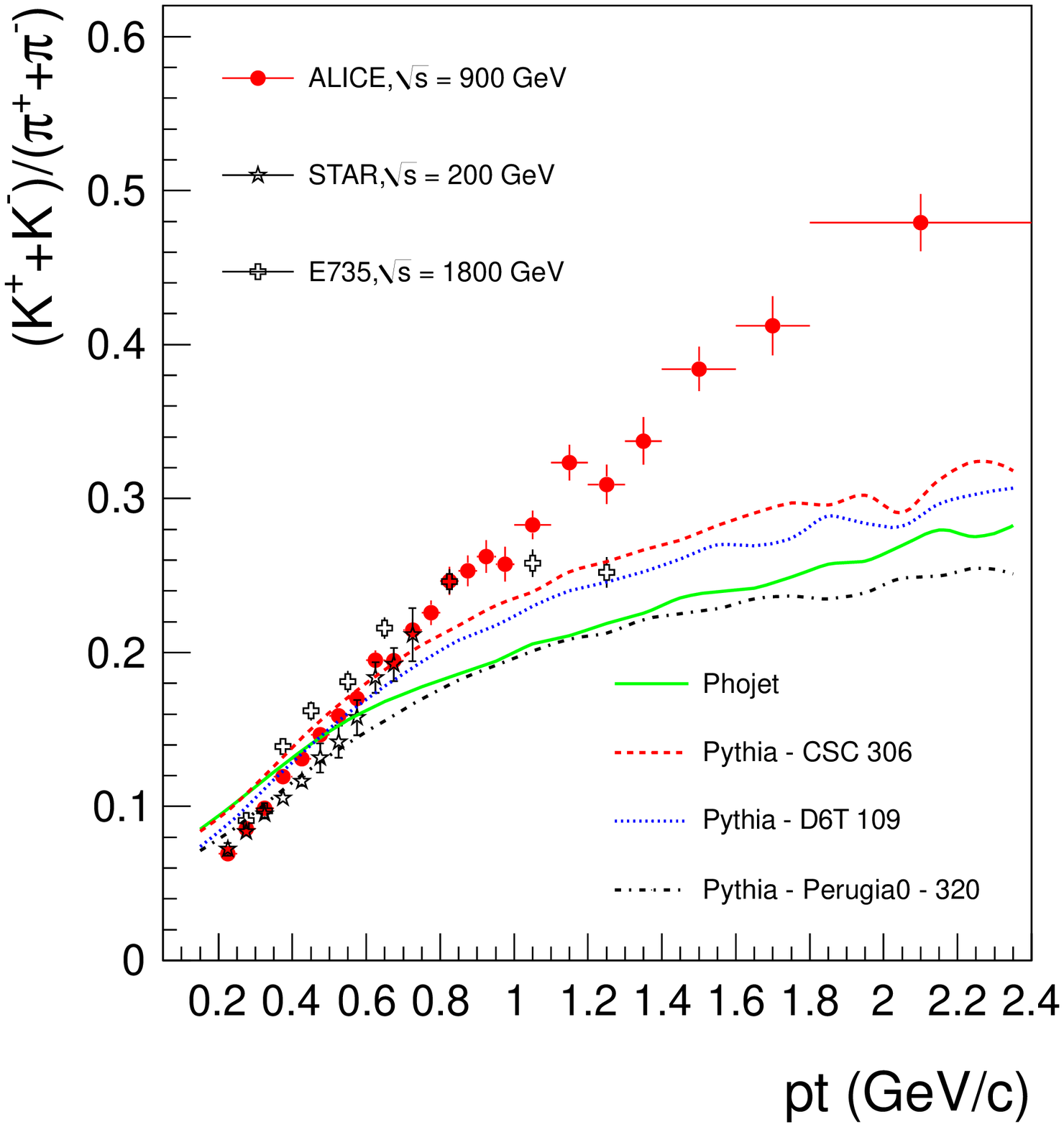}
\caption{\label{fig:ratios1}  Ratio of (\kap + \kam)/(\pip + \pim)
as a function of  \ensuremath{p_{\rm t}} from pp collisions at \s = 900  GeV  (statistical
errors only). Values from the E735
Collaboration~\cite{Alexopoulos:1993wt} and the STAR
Collaboration also are given. The
dashed and dotted curves refer to calculations using PYTHIA and
PHOJET at \s = 900  GeV.}
\end{minipage}\hspace{2pc}%
\begin{minipage}{16pc}
\includegraphics[width=15pc]{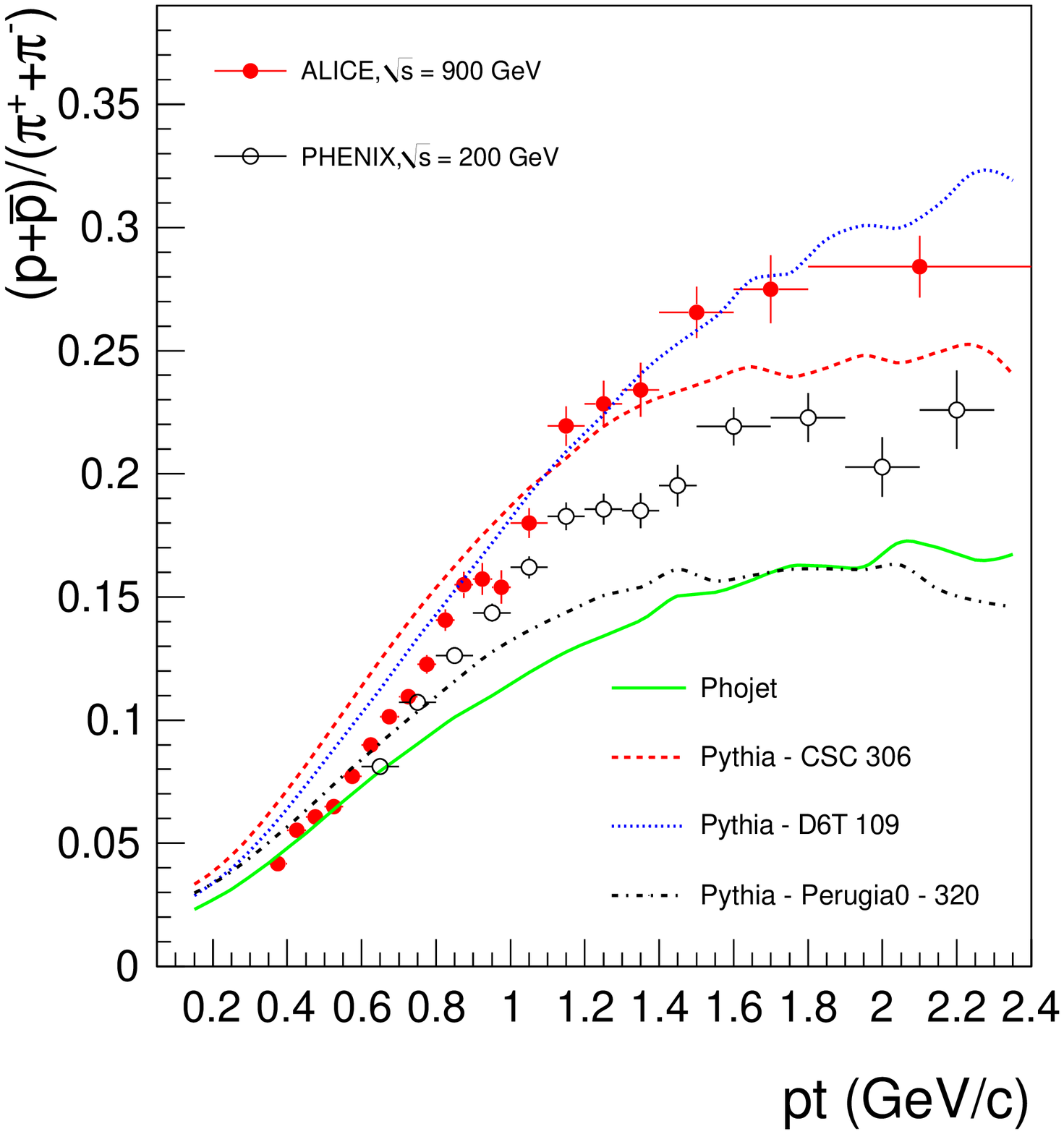}
\caption{\label{fig:ratios2} Ratio (p + \pbar) / (\pip + \pim) as a
function of  \ensuremath{p_{\rm t}} from pp collisions at \s = 900  GeV  (statistical
errors only). Values from the PHENIX
Collaboration~\cite{Adler:2006xd} also are given. The
dashed and dotted curves refer to calculations using PYTHIA and
PHOJET at \s = 900  GeV. }
\end{minipage} 
\end{figure}

Figure~\ref{fig:ratios1} shows the
\ensuremath{p_{\rm t}}-depen\-dence of the (\kap + \kam)/(\pip + \pim) ratio and also the measurements by the
E735~\cite{Alexopoulos:1993wt} and STAR
Collaborations~\cite{:2008ez}. It can be seen that the observed
increase of K/$\pi$ with \ensuremath{p_{\rm t}} does not depend strongly on
collision energy.

A comparison with event generators shows that at \ensuremath{p_{\rm t}} $>$ 1.2
 GeV/c, the measured K/$\pi$ ratio is larger than any of the
model predictions.

In Figure~\ref{fig:ratios2}, the measured
(p + \pbar) / (\pip + \pim) ratio is compared to results at \s = 200  GeV  from the
PHENIX Collaboration~\cite{Adler:2006xd}. Both measurements are
feed-down corrected. At low \ensuremath{p_{\rm t}}, there is no energy-dependence of
the p$/\pi$ ratio visible, while at higher \ensuremath{p_{\rm t}}$ > 1$ ~GeV/c, the
p/$\pi$ ratio is larger at \s = 900  GeV  than at \s = 200  GeV 
energy.

Event generators seem to separate into two groups, one with high
p/$\pi$ ratio (PYTHIA CSC and D6T), which agree better with the
data and one group with a lower p/$\pi$ ratio (PHOJET and PYTHIA
Perugia0), which are clearly below the measured values. These
comparisons can be used for future tunes of the event generators.

\begin{figure}[h]
\includegraphics[width=16pc]{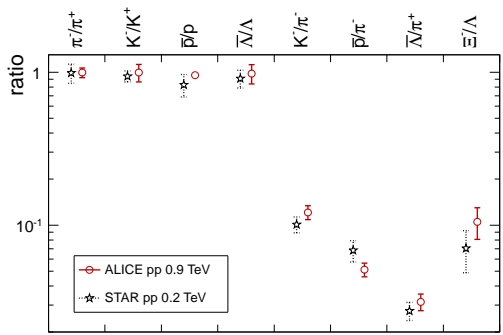}\hspace{2pc}%
\begin{minipage}[b]{16pc}\caption{\label{fig:asrat}Ratios of integrated yields including $\pi^{(\pm)}$, K$^{(\pm)}$, p and \pbar~ 
performed with the ALICE experiment~\cite{Alice:Pid,Aamodt:2010dx} and compared
with STAR values for pp collisions at $\sqrt{s}$~=~0.2 TeV \cite{Abelev:2006cs}.
All ratios are feed-down corrected.
Statistical and systematic uncertainties are added in quadrature.}
\end{minipage}
\end{figure}

Using the particle integrated yields, a comparison with STAR feed-down corrected particle
ratios at $\sqrt{s}$~=~0.2 TeV \cite{Abelev:2006cs} is shown in Fig.~\ref{fig:asrat}.
With the centre of mass energy increasing from $\sqrt{s}$~=~0.2 TeV to 0.9 TeV the measured ratios are similar except the $\bar{p}/\pi^{-}$ ratio which decreases
slightly from $0.068 \pm 0.011$ to $0.051 \pm 0.005$.
The strange to non-strange particle ratios seem to increase but stay compatible
within uncertainties:
the $\rm{K}^{-}/\pi^{-}$ from $0.101 \pm 0.012$ to $0.121 \pm 0.013$
and the \lam$/\pi^{+}$ from $0.027 \pm 0.004$ to $0.032 \pm 0.003$.
  
\begin{figure}[h]
\includegraphics[width=16pc]{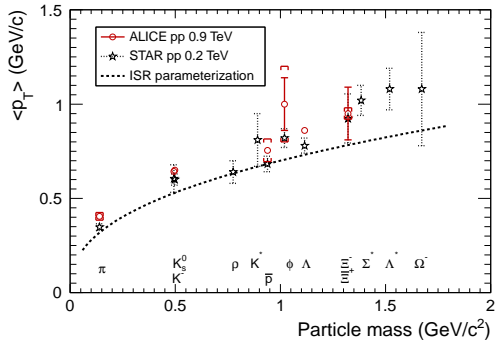}\hspace{2pc}%
\begin{minipage}[b]{16pc}\caption{\label{fig:meanpt}$\ensuremath{\langle p_{\mathrm{t}}} \rangle$ vs. particle mass for the measurements performed with the ALICE experiment
and compared with STAR values for pp collisions at $\sqrt{s}$~=~0.2 TeV \cite{Abelev:2006cs,Adams:2006yu}
and the ISR parameterization~\cite{Bourquin:1976fe}.
Both statistical (vertical error bars) and systematic (brackets) uncertainties are shown for ALICE data. }
\end{minipage}
\end{figure}

The yields and $\ensuremath{\langle p_{\mathrm{t}}} \rangle$ obtained with the ALICE experiment are compared for each
particle with existing data at the same energy and also with results at
lower and higher energies.
The various experiments differ in acceptance and event selection (i.e. NSD or INEL) but
the dependence of $\ensuremath{\langle p_{\mathrm{t}}} \rangle$ with respect to these variables is found to be negligible.
Consequently the $\ensuremath{\langle p_{\mathrm{t}}} \rangle$ values are directly comparable, whereas the comparison
of the yields can require further scaling because of different (pseudo)rapidity coverages. 
Figure~\ref{fig:meanpt} reports $\ensuremath{\langle p_{\mathrm{t}}} \rangle$ measurements along with those of the
STAR experiment~\cite{Abelev:2006cs,Adams:2006yu}.
It is remarkable that the $\ensuremath{\langle p_{\mathrm{t}}} \rangle$ remains close to the ISR parameterization~\cite{Bourquin:1976fe}
although the collision energy increased by a factor $36$.
  
\section{Conclusions}
The first analysis of transverse
momentum spectra of identified hadrons,  \pip, \pim, \kap, \kam,
p, \pbar, \kos, \lam, \lambar, \xip, \xim and $\phi$ in pp collisions at \s\ = 900  GeV  with the ALICE
detector is presented.
Various particle identification techniques have been used and this allows us to cover a broad momentum range.
Agreement in the K/$\pi$ ratio is seen when comparing to \pbar p
collisions at the Tevatron and Sp\pbar S. Comparing our results
with similar measurement from the STAR Collaboration using pp
collisions at \s = 200  GeV  the shape of the spectra shows an
increase of the hard component, but we observe only a slight
increase of the mean-\ensuremath{p_{\rm t}} values.

The integrated yields and average transverse momenta of mesons containing strange quarks (\kos~and $\phi$)
and hyperons (\lam, \lambar~ and $\Xi$) have been compared with earlier data collected in pp and p\pbar~ interactions at various energies.
These results provide a useful baseline for comparisons with recent tunes
of the PYTHIA model and a reference for future measurements in heavy-ion
collisions at the LHC.

These studies demonstrate the precision with which ALICE can measure charged hadrons, resonances and topologically reconstructed weakly decaying particles.
Measurements of these particles is a substantial part of the ALICE program in both pp and PbPb collisions.


\end{document}